# Transfer and teleportation of system-environment entanglement

Tytus Harlender
*Department of Theoretical Physics, Faculty of Fundamental Problems of Technology,
Wrocław University of Science and Technology, 50-370 Wrocław, Poland*

Katarzyna Roszak
*Department of Theoretical Physics, Faculty of Fundamental Problems of Technology,
Wrocław University of Science and Technology, 50-370 Wrocław, Poland*

(Dated: July 1, 2021)

We study bidirectional teleportation while explicitly taking into account an environment. This environment initially causes pure dephasing decoherence of the Bell state which assists teleportation. We find that when teleportation is performed in one direction it is accompanied by a transfer of correlations into the post-teleportation state of qubit *C*, which results in decoherence of the state. In the other direction, if no new decoherence process occurs, we find that not only the state of the qubit but also its correlations with an environment are being teleported with unit Fidelity. These processes do not depend on the measurement outcome during telportation and do not differentiate between classical and quantum correlations. If, on the other hand, the second teleportation step is preceded by decoherence of the Bell state then the situation is much more complicated. Teleportation and transfer of correlations occur simultaneously, yielding different teleported qubit-environment states for different measurement outcomes. These states can differ in the degree of coherence of the teleported qubit, but only for an entangling Bell-state-environment interaction in the first step of teleportation, can they have different amounts of qubit-environment entanglement. In the extreme case, one of the teleported qubit states can be entangled with the environment while the other is separable.

## I. INTRODUCTION

Teleportation [1] is a pivotal example of the consequences and importance of entanglement. Using a two-qubit entangled state one can transport an unknown qubit state onto another qubit. Since fundamentally in quantum mechanics particles are indistinguishable, this post-teleportation qubit is indistinguishable from the pre-teleportation one by any measurement. This means that it is the same qubit. It has been also shown that teleportation can be viewed as a primitive subroutine of quantum computation [2].

Since its original discovery [1], teleportation has been extensively studied theoretically. Firstly, more complex teleportation scenarios have been devised [3–5], allowing for teleportation of quantum states of larger ensembles using larger maximally entangled states. Furthermore, the effects of teleportation via a non-maximally entangled state on the Fidelity of teleportation have received much attention [3, 6–13]. Here, a particularly important case is related to the loss of coherence [3, 9–13], since decoherence is rarely avoidable in realistic qubit realizations. It has been found that when the average teleportation Fidelity over all possible qubit states is smaller than 2/3, the bipartite state used to perform teleportation must be separable [1, 11, 14].

The first experimental realizations of teleportation were performed quite soon after the theoretical prediction, most notably on optical systems [15–17], but there were also successful experiments performed on NMR [18]. Later realizations on different systems have been demonstrated, such as atoms, ions, and particles [19, 20], teleportation between light and matter [21], teleportation of continous variables [22, 23], etc. Simultanously the distances over which successful teleportation has been demonstrated have been extended until a definitely macroscopic range of teleportation has been obtained [24–26].

Recently, interesting effects resulting from more realistic theoretical treatment of the processes which lead to decoherence have been found. For example, it has been shown that local noise can enhance two-qubit teleportation while it does not increase entanglement of the teleported state [27]. Furthermore, non-local memory effects allow for teleportation Fidelity enhancement when teleporting using a mixed entangled state [28]. These results suggest that the nature of the interaction with the environment is important when performing teleportation and the environment could possibly be used to assist faithful teleportation.

We study bidirectional teleportation [29, 30] in the simplest possible scenario when an unknown qubit state is teleported via a Bell state. We include an interaction between the Bell state and an environment and keep the degrees of freedom of the environment explicitly while performing teleportation on the qubits. This allows us to study not only the effect that the environment has on the qubit states during the procedure, but also the effect that the qubits have on the environment. We are especially interested in the behavior of correlations of specific parts of the three-qubit system and the environment, whether quantum or classical.

We find that in the first step of teleportation, the correlations which are initially present between the two-qubit Bell state and the environment are transferred into one qubit while information about the unknown qubit state is teleported. This yields a qubit-environment state in which all of the information about the unknown state is present, but the correlations with the environment make it impossible to read it out from the qubit state alone.

If the teleportation is performed in the other direction without additional decoherence then we observe perfect teleportation of the qubit-environment state. Hence, not only the state of the qubit (which is mixed) is teleported with unit Fidelity, but the correlations which were present between qubit $C$ and the environment have been teleported to qubit $A$. This is teleportation rather than transfer, since no correlations with the environment were ever previously present for qubit $A$. Note that there is an infinite number of qubit-environment density matrices which yield the same mixed qubit state, so the perfect conveyance of the correlations is by no means obvious.

For the two processes described above, the measurement outcome in the teleportation procedure is irrelevant. Furthermore, there is no quantitative or qualitative difference between the situation when the interaction with the environment leads to entanglement or not.

Once the two processes occur simultaneously when we allow new decoherence to take place between teleportations, there is a qualitative change in the results. Firstly, the measurement outcome is no longer irrelevant and there are two disting qubit-environment states which can be obtained after teleportation. They can differ in qubit coherence, but if the decoherence in the first step of teleportation is entangling then they can also differ in the amount of qubit-environment entanglement. This difference can be arbitrarily large and it is possible for one state to be separable while the other is entangled.

The paper is organized as followed. The basic concept of teleportation is recounted in Sec. II. Teleportation in one direction with the help of a decohered Bell state is studied in Sec. III, while teleportation in the other direction when no additional decoherence process has occured is studied in Sec. IV. In Sec. V we study the same teleportation process as in Sec. IV, but for the situation when the second teleportation procedure is precluded by decoherence of the Bell state. Examples for Sec. IV are provided in Sec. VI. Sec. VII concludes the paper.

## II. TELEPORTATION

The protocol of teleportation of an unknown qubit state [1], regardless of its huge implications for the importance and consequences of entanglement, is particularly straightforward. Three qubits are necessary: Qubit $A$, the unknown quantum state of which is to be teleported,

$$|\psi\rangle_A = \alpha|0\rangle_A + \beta|1\rangle_A, \tag{1}$$

and two qubits $B$ and $C$, which are initially in a maximally entangled state. Here we will assume that this state is a chosen Bell state,

$$|\Phi_+\rangle_{BC} = \frac{1}{\sqrt{2}}\left(|00\rangle_{BC} + |11\rangle_{BC}\right). \tag{2}$$

The procedure itself involves a measurement of qubits $A$ and $B$ in the Bell basis which transfers the information about the coefficients $\alpha$ and $\beta$ into the state of qubit $C$, followed by a unitary operation on qubit $C$, which is dependent on the outcome of the measurement. The unitary operation transforms the post-measurement state of qubit $C$ into state (1). Note that this procedure faithfully teleports the state from qubit $A$ to qubt $C$, even if the initial state of qubit $A$ is mixed.

In the following, we will implicitly assume that the measurement outcome on qubits $A$ and $B$ is the Bell state (2). This is because this measurement outcome guarantees the teleportation from qubit $A$ to $C$ without the necessity of performing an additional unitary operation. This choice bears no conceptual consequence on the results presented in the next two sections, as what is obtained is equivalent regardless of the outcome, so it is made strictly for convenience.

## III. TELEPORTATION WITH THE HELP OF DECOHERED BELL STATE

Teleportation by means of a non-maximally entangled state has already been extensively studied, both in the case of pure states [6–8], and in the situation when the decrease in entanglement is related to some form of decoherence [3, 9–13]. We will study the situation when the Bell state (2) undergoes pure dephasing due to an interaction with an environment. What is special in our approach is that we will not trace out the environment and instead study the full density matrix of the three qubits and the environment. This allows us to keep track of what happens with the correlations that have formed throughout the evolution during the teleportation procedure and identify which processes are accompanied by transfer or teleportation of correlations and how faithful they are.

To this end we assume that between initialization of the three qubits in state $|\psi\rangle_A \otimes |\Phi_+\rangle_{BC}$ and the measurement of qubits $A$ nd $B$ in the Bell basis, there is a time $\tau$ during which qubits $B$ and $C$ interact with an environment (E) according to a pure dephasing Hamiltonian [31],

$$\hat{H}_{PD} = \sum_{i,j=0,1} |ij\rangle_{BCBC}\langle ij| \otimes \hat{V}_{ij}. \tag{3}$$

Here the operators $\hat{V}_{ij}$ act on the subspace of the environment. The evolution operator corresponding to Hamiltonian (3) is given by [31]

$$\hat{U}_{PD}(\tau) = \sum_{i,j=0,1} |ij\rangle_{BCBC}\langle ij| \otimes \hat{w}_{ij}(\tau), \tag{4}$$

with

$$\hat{w}_{ij}(\tau) = \exp\left(-\frac{i}{\hbar}\hat{V}_{ij}\tau\right) \tag{5}$$

We assume that initially the environment is in a product state with the qubits, and the full initial state of system $ABCE$ is given by

$$\hat{\sigma}(0) = |\psi\rangle_{AA}\langle\psi| \otimes |\Phi_+\rangle_{BCBC}\langle\Phi_+| \otimes \hat{R}(0), \tag{6}$$

where $\hat{R}(0)$ is the initial state of the environment. We do not make any assumptions on the environmetal density matrix. Hence, using the evolution operator (4) we find the state of the qubits and the environment at time $\tau$,

$$\hat{\sigma}(\tau) = |\psi\rangle_{AA}\langle\psi| \otimes \frac{1}{2}\begin{pmatrix} \hat{R}_{00}(\tau) & 0 & 0 & \hat{R}_{01}(\tau) \\ 0 & 0 & 0 & 0 \\ 0 & 0 & 0 & 0 \\ \hat{R}_{10}(\tau) & 0 & 0 & \hat{R}_{11}(\tau) \end{pmatrix}_{BCE}. \tag{7}$$

In eq. (7) the density matrix on the right of the tensor product describes the joint state of qubits $B$, $C$ and the environment. It is written in a notation which is convenient for interactions leading to pure dephasing [32, 33], where the matrix form is used with respect to the pointer basis of the two qubits, $|ij\rangle_{BC}$, while $\hat{R}_{ij}(\tau)$ are environmental operators with

$$\hat{R}_{ij}(\tau) = \hat{w}_{ii}(\tau)\hat{R}(0)\hat{w}_{jj}^\dagger(\tau). \quad (8)$$

Since $\hat{R}_{00}(\tau)$ and $\hat{R}_{11}(\tau)$ are density matrices, tracing out the environmental degrees of freedom from the matrix (7) yields a dephased Bell state,

$$\hat{\rho}_{BC}(\tau) = \frac{1}{2}\begin{pmatrix} 1 & 0 & 0 & c(\tau) \\ 0 & 0 & 0 & 0 \\ 0 & 0 & 0 & 0 \\ c^*(\tau) & 0 & 0 & 1 \end{pmatrix}, \quad (9)$$

with $c(\tau) = \text{Tr}_E \hat{R}_{01}(\tau)$.

We now perform a projective measurement in the Bell basis on qubits $A$ and $B$ for the whole, three qubit and environment, system in state (7). If the outcome is (2), then the post-measurement state is

$$\hat{\sigma}_{PM}(\tau) = |\Phi_+\rangle_{ABAB}\langle\Phi_+| \otimes \begin{pmatrix} |\alpha|^2 \hat{R}_{00}(\tau) & \alpha^*\beta \hat{R}_{01}(\tau) \\ \alpha\beta^* \hat{R}_{10}(\tau) & |\beta|^2 \hat{R}_{11}(\tau) \end{pmatrix}_{CE}. \quad (10)$$

When the post measurement state of qubit $C$ and the environment on the right side of the tensor product in eq. (10) is compared to the pre measurement $BCE$ state [on the right side of the tensor product in eq. (7)], it is apparent that two processes took place. On one hand, the coefficients of the qubit $A$ state (1) have been teleported to qubit $C$, but simultanously the correlations with the environment that were present with qubits $B$ and $C$ have been fully transferred to qubit $C$.

These correlations are the reason that the teleported state of qubit $C$ is dephased. In fact, the degree of coherence of qubit $C$ is the same as the previous degree of coherence of the Bell state, since the state of qubit $C$ is now given by

$$\hat{\rho}_C^{PM}(\tau) = \begin{pmatrix} |\alpha|^2 & \alpha^*\beta c(\tau) \\ \alpha\beta^* c^*(\tau) & |\beta|^2 \end{pmatrix}, \quad (11)$$

with the same dephasing coefficient $c(\tau)$ as in eq. (9).

The system-environment correlations which are present in eqs (7) and (10) may either be quantum (with an entangled system-environment state) or classical (with a separable system-environment state) if the initial state of the environment is mixed [31, 34]. It is important to note here that the nature of the correlations cannot change during teleportation.

The qubit-environment entanglement measure tailored to pure dephasing evolutions, which was introduced in Ref. [35], can be used to quantify the amount of entanglement in the CE state in eq. (10). It can also be used to quantify entanlgement of the BCE state of eq. (7) because during pure dephasing the state of qubits $B$ and $C$ is effectively confined to a two-dimensional subspace. The measure is given by $E_{BCE} = \left[1 - F\left(\hat{R}_{00}(\tau), \hat{R}_{11}(\tau)\right)\right]$ for the dephased Bell state and by

$$E_{CE} = 4|\alpha|^2|\beta|^2 \left[1 - F\left(\hat{R}_{00}(\tau), \hat{R}_{11}(\tau)\right)\right] = 4|\alpha|^2|\beta|^2 E_{BCE} \quad (12)$$

for the post-teleportation state. Here the function $F(\hat{\rho}_1, \hat{\rho}_2) = \left[\text{Tr}\sqrt{\sqrt{\rho_1}\rho_2\sqrt{\rho_1}}\right]^2$ denotes the Fidelity. As seen from eq. (12) the amount of entanglement transferred depends strongly on the teleported state (1), but teleportation retains the nature of the correlation during transfer. If the state of qubit $A$ before teleportation is either $|0\rangle$ or $|1\rangle$, entanglement will not be transferred, but these are also the only states which will be teleported faithfully. Any superposition state will acquire a part of entanglement with the environment if it is present in state (7) and this fraction is given by $4|\alpha|^2|\beta|^2$. If, on the other hand, the decoherence of the Bell state is separable in its nature then no quantum correlations can be present in the post teleportation state (qubit-environment entanglement for pure dephasing evolutions is equivalent to the quantum discord [36]).

## IV. TELEPORTATION OF DECOHERED STATE

We will now study teleportation from qubit $C$ to qubit $A$ assuming no delay time between the procedure described in the previous section. Hence, the pre-teleportation state is given by eq. (10), so qubit $C$ is correlated with the environment while qubits $A$ and $B$ are in Bell state (2). The two-qubit measurement is performed on qubits $B$ and $C$ and we again assume that the outcome is the state (2), for clarity.

After teleportation, the system of three qubits and environment is given by

$$\hat{\sigma}_{PM2}(\tau) = \begin{pmatrix} |\alpha|^2 \hat{R}_{00}(\tau) & \alpha^*\beta \hat{R}_{01}(\tau) \\ \alpha\beta^* \hat{R}_{10}(\tau) & |\beta|^2 \hat{R}_{11}(\tau) \end{pmatrix}_{AE} \otimes |\Phi_+\rangle_{BCBC}\langle\Phi_+|. \quad (13)$$

Comparing this state with the pre-teleportation state (10), we note that the state of qubit $A$ and the environment is exactly the same as the state of qubit $C$ and environment was before teleportation. If the environmental degrees of freedom are traced out of the post-teleportation $AE$ state, we will obtain exactly the pre-teleportation state of qubit $C$, which is given by eq. (11). It is relevant to note here that there is an infinite number of $AE$ states that yield the same dephased state of qubit $A$. For a mixed environment, such $AE$ states can differ substantially, as there exist both entangled and separable states that lead the same amount of qubit decoherence. The states are not equivalent as entanglement easily manifests itself in e. g. post-measurement qubit evolution [37–39], so the exact transfer of correlations is a relevant factor.

Hence, not only the state of qubit $C$ was teleported with unit Fidelity to qubit $A$, but the state of qubit $A$ and its environment is now exactly the same as the initial state of qubit $C$ and the same environment. Qubit-environment correlations have been faithfully teleported during the process.

## V. SIMULTANEOUS TRANSFER AND TELEPORTATION OF CORRELATIONS

In this section we would like to study the situation when the teleportation back from qubit $C$ to qubit $A$ is preceded by a decohering of the Bell state, similarly as in Sec. III. Hence we start with the *ABCE* state (10) and allow qubits *A* and *B* to undergo pure dephasing for time $t$. The process is governed by a Hamiltonian with the same structure as Hamiltonian (3), hence the evolution operator (4) and conditional evolution operators (5) also retain the structure. We do not assume that it is the same Hamiltonian, so any operators pertaining to this second process will be labeled with a prime.

Since there are correlations between qubit *C* and the environment already present in state (10), the newly decohered state cannot be written in product form between parts *AB* and *CE* as eq. (10), nor in a product form between parts *ABE* and *C*, equivalently to eq. (7). In most situations there are at least classical correlations present in the partitions. The state is given by

$$\hat{\sigma}'_{PM}(\tau,t) = \qquad\qquad\qquad (14)$$
$$\frac{1}{2}\begin{pmatrix} |\alpha|^2\hat{R}^{00}_{00}(\tau,t) & \alpha^*\beta\hat{R}^{00}_{01}(\tau,t) & |\alpha|^2\hat{R}^{01}_{00}(\tau,t) & \alpha^*\beta\hat{R}^{01}_{01}(\tau,t) \\ \alpha\beta^*\hat{R}^{00}_{10}(\tau,t) & |\beta|^2\hat{R}^{00}_{11}(\tau,t) & \alpha\beta^*\hat{R}^{01}_{10}(\tau,t) & |\beta|^2\hat{R}^{01}_{11}(\tau,t) \\ |\alpha|^2\hat{R}^{10}_{00}(\tau,t) & \alpha^*\beta\hat{R}^{10}_{01}(\tau,t) & |\alpha|^2\hat{R}^{11}_{00}(\tau,t) & \alpha^*\beta\hat{R}^{11}_{01}(\tau,t) \\ \alpha\beta^*\hat{R}^{10}_{10}(\tau,t) & |\beta|^2\hat{R}^{10}_{11}(\tau,t) & \alpha\beta^*\hat{R}^{11}_{10}(\tau,t) & |\beta|^2\hat{R}^{11}_{11}(\tau,t) \end{pmatrix},$$

where the sixteen environmental matrices are obtained from eq. (8) following

$$\hat{R}^{kq}_{ij}(\tau,t) = \hat{w}'_{kk}(t)\hat{R}_{ij}(\tau)\hat{w}'^{\dagger}_{qq}(t). \qquad (15)$$

The matrix form in eq. (14) corresponds to the states of the three qubits, where the basis is arranged in the following order: $\{|000\rangle,|001\rangle,|110\rangle,|111\rangle\}$. The other four elements of the three-qubit basis have been omitted, as all other matrix elements are equal to zero.

Let us now teleport the state of qubit *C* to qubit *A*. Contrarily to the results of Secs III and IV, the measurement outcome on qubits *B* and *C* actually matters here. Regardless of the measurement outcome $|\lambda\rangle_{BC}$, the post-teleportation state (after the measurement and apropriate unitary transformation on qubit *A*) will be of the form

$$\hat{\sigma}'_{PM2}(\tau,t) = \hat{\rho}^{\lambda}_{AE}(\tau,t) \otimes |\lambda\rangle_{BC\,BC}\langle\lambda|, \qquad (16)$$

but the state of qubit *A* and environment will be given by

$$\hat{\rho}^{\Phi_\pm}_{AE}(\tau,t) = \begin{pmatrix} |\alpha|^2\hat{R}^{00}_{00}(\tau,t) & \alpha^*\beta\hat{R}^{01}_{01}(\tau,t) \\ \alpha\beta^*\hat{R}^{10}_{10}(\tau,t) & |\beta|^2\hat{R}^{11}_{11}(\tau,t) \end{pmatrix}_{AE}, \qquad (17)$$

for either outcome $|\Phi_\pm\rangle = (|00\rangle \pm |11\rangle)/\sqrt{2}$, and we get

$$\hat{\rho}^{\Psi_\pm}_{AE}(\tau,t) = \begin{pmatrix} |\alpha|^2\hat{R}^{11}_{00}(\tau,t) & \alpha^*\beta\hat{R}^{10}_{01}(\tau,t) \\ \alpha\beta^*\hat{R}^{01}_{10}(\tau,t) & |\beta|^2\hat{R}^{00}_{11}(\tau,t) \end{pmatrix}_{AE}, \qquad (18)$$

for either outcome $|\Psi_\pm\rangle = (|01\rangle \pm |10\rangle)/\sqrt{2}$. The states (17) and (18) are obviously different, and although for both states the initial correlations of qubit *C* with the environment were teleported faithfully [as indicated by the subscript in the matrices $\hat{R}^{kq}_{ij}(\tau,t)$, which always correspond to the matrices in eq. (10)], different correlations from state (14) are transported into the state (the superscript pertain to the later decoherence process).

Both states (17) and (18) yield a purely dephased state of qubit *A* after the environment is traced out, but the degree of coherence can be different. The density matrices describing the state of qubit *C* alone retain the form of eq. (11), but with

$$c^{\Phi_\pm}(\tau,t) = \mathrm{Tr}\left(\hat{w}'^{\dagger}_{11}(t)\hat{w}'_{00}(t)\hat{R}_{01}(\tau)\right), \qquad (19\mathrm{a})$$

$$c^{\Psi_\pm}(\tau,t) = \mathrm{Tr}\left(\hat{w}'^{\dagger}_{00}(t)\hat{w}'_{11}(t)\hat{R}_{01}(\tau)\right). \qquad (19\mathrm{b})$$

The two quantities are obviously the same if the two conditional evolution operators of the later decoherence process are Hermitian and commute, $[\hat{w}'_{00}(t),\hat{w}'_{11}(t)] = 0$, $\hat{w}'_{ii}(t) = \hat{w}'^{\dagger}_{ii}(t)$. They are complex conjugates of each other if the matrix $\hat{R}_{01}(\tau)$ is Hermitian.

If there is entanglement in either *AE* state (17) or (18), it is of the type which can be qualified by the condition of Ref. [34], so the if and only if conditions of separability for states (17) and (18) are

$$\hat{w}'_{00}(t)\hat{R}_{00}(\tau)\hat{w}'^{\dagger}_{00}(t) = \hat{w}'_{11}(t)\hat{R}_{11}(\tau)\hat{w}'^{\dagger}_{11}(t), \qquad (20\mathrm{a})$$

$$\hat{w}'_{00}(t)\hat{R}_{11}(\tau)\hat{w}'^{\dagger}_{00}(t) = \hat{w}'_{11}(t)\hat{R}_{00}(\tau)\hat{w}'^{\dagger}_{11}(t), \qquad (20\mathrm{b})$$

respectively, while the condition of separability of qubit *C* and the environment pre-teleportation and decoherence (10) is $\hat{R}_{00}(\tau) = \hat{R}_{11}(\tau)$. Hence, if there is no entanglement after the first part of teleportation then both states are either entangled or separable, and the amount of entanglement in the two states is also the same [35], as it is quantified by a function analogous to eq. (12),

$$E_{CE}(\tau,t) = 4|\alpha|^2|\beta|^2\left[1 - F\left(\hat{R}^{kk}_{00}(\tau,t),\hat{R}^{qq}_{11}(\tau,t)\right)\right], \qquad (21)$$

with $k = 0$ and $q = 1$ for state (17) and $k = 1$ and $q = 0$ for state (18). Since for a separable state (10), we have $\hat{R}^{00}_{00}(\tau,t) = \hat{R}^{00}_{11}(\tau,t)$ and $\hat{R}^{11}_{00}(\tau,t) = \hat{R}^{11}_{11}(\tau,t)$, the function (21) yields the same outcome for both states. Note, that this does not necessarily translate to the same degree of coherence, as discussed in the previous paragraph.

When there is entanglement between qubit *C* and the environment in state (10) then the conditions of separability for states (17) and (18) differ from each other. The two states can, in this case, not only have a different amount of qubit-environment entanglement, but the situation when one of the states is separable while the other is entangled can be realized. This is a direct consequence of the type of entanglement between the three qubits and the environment present in state (14). The conditions (20) constitute two of the seven nontrivial separability conditions of the first type for an 8-dimensional system and an environment interacting via a pure-dephasing Hamiltonian [31], which are applicable in case of a density matrix of the form (14). Since this is effectively a $4 \times 4$ system, then a further four conditions are automatically fulfilled, and only one nontrivial separability condition of this type is irrelevant in case of the teleportation process under study. None of the separability conditions of the second type [31] are relevant here, because we are dealing with transfer of entanglement to a single qubit and the second type of entanglement has no single-qubit equivalent.

## VI. EXAMPLE: A SINGLE QUBIT ENVIRONMENT

To illustrate the results of the previous section we will study an exemplary system environment evolution with the smallest possible environment, one composed of a single qubit. We will assume that the initial state of the environment is $\hat{R}(0) = c_0|0\rangle\langle 0| + c_1|1\rangle\langle 1|$, so it is pure for $c_0 = 0, 1$ and maximally mixed for $c_0 = c_1 = 1/2$. We will further assume, for simplicity, that the interaction with the environment is fully asymmetric in both processes, meaning that $\hat{w}_{11}(\tau) = \hat{w}'_{11}(t) = \mathbb{I}$.

Let us look at the situation when the first interaction, the one described in Sec. III, yields a separable qubit-environment state at all times $\tau$. This means that the operator $\hat{w}_{00}(\tau)$ must be diagonal in the same basis as the initial state of the environment at all times [34], so it can be written as $\hat{w}_{00}(\tau) = e^{i\varphi_0\tau}|0\rangle\langle 0| + e^{i\varphi_1\tau}|1\rangle\langle 1|$, and we get $\hat{R}_{01}(\tau) = c_0 e^{i\varphi_0\tau}|0\rangle\langle 0| + c_1 e^{i\varphi_1\tau}|1\rangle\langle 1|$. We will write the conditional evolution operator governing the second decoherence process in a general way,

$$\hat{w}'_{00}(t) = e^{i\phi_a t}|a\rangle\langle a| + e^{i\phi_b t}|b\rangle\langle b|. \quad (22)$$

Here $|a\rangle = x|0\rangle + y|1\rangle$ and $|b\rangle = y^*|0\rangle - x^*|1\rangle$ are two orthogonal states that diagonalize the respective part of the Hamiltonian. The parameters $x$ and $y$, $|x|^2 + |y|^2 = 1$, depend on the interaction Hamiltonian and are used as parameters in this example.

It is now straightforward to find the values of the degree of coherence factors (19), which are given by

$$c^\lambda(\tau, t) = |x|^2 \left[ c_0 e^{i(\varphi_0\tau \pm \phi_a t)} + c_1 e^{i(\varphi_1\tau \pm \phi_b t)} \right] \quad (23)$$
$$+ |y|^2 \left[ c_0 e^{i(\varphi_0\tau \pm \phi_b t)} + c_1 e^{i(\varphi_1\tau \pm \phi_a t)} \right],$$

with pluses for $\lambda = \Phi_\pm$ and minuses for $\lambda = \Psi_\pm$. In Fig. 1 we plot the absolute value of $c^\lambda(\tau, t)$ in both cases for a maximally mixed environment (so both qubit-environment states are always separable) for $\varphi_0 = 0$ and $\phi_a = 0$, and for different values of $|x|^2$. We set the first decoherence time so that $\varphi_1\tau = \pi/2$ and the state teleported in the first part of the procedure is dephased, and show the degree of coherence as a function of $\phi_b t$. For $|x|^2 = 0.5$ both curves are the same, but in the other two cases, the degree of coherence of qubit $A$ after the second teleportation strongly depends on the measurement outcome on qubits $B$ and $C$.

We will now look at a situation when both interactions can lead to entanglement to illustrate the fact that depending on the measurement result in the second teleportation process we can have a qubit which is entangled with its environment or not. To this end, let us assume that both operators $\hat{w}_{00}(\tau)$ and $\hat{w}'_{00}(t)$ are the same and are given by eq. (22) with $t = \tau$, while the other two conditional evolution operators remain trivial. We then easily find that $\hat{R}^{11}_{00}(t,t) = \hat{R}^{00}_{11}(t,t) = \hat{w}_{00}(t)\hat{R}(0)\hat{w}^\dagger_{00}(t)$, so the separability condition (20b) is fulfilled for all $t$ and there is not qubit-environment entanglement in state (18).

The two environmental operators relevant for entanglement in state (17) are given by $\hat{R}^{00}_{00}(t,t) = \hat{w}_{00}(2t)\hat{R}(0)\hat{w}^\dagger_{00}(2t)$ and $\hat{R}^{00}_{11}(t,t) = \hat{R}(0)$, so the separability condition (20a) is not fulfilled unless $x = 0$ or $y = 0$ outside of discrete points in time. Entanglement measured by the function (21) for $x = y = 1/\sqrt{2}$ is plotted in Fig. 2 for different initial mixedness of the environment and for an equal superposition teleported state (1).

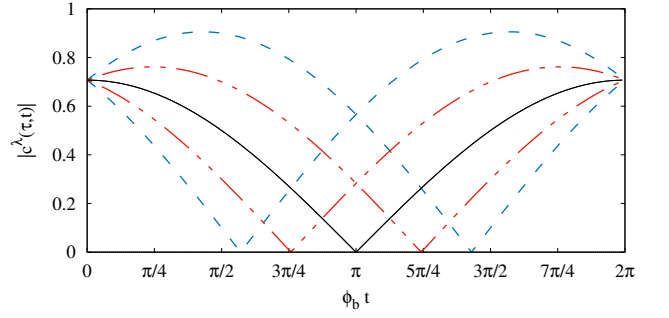

FIG. 1. Exemplary degree of coherence (absolute value) for $\lambda = \Phi_\pm$ and $\lambda = \Psi_\pm$ for a single qubit environment initially in a maximally mixed state, when no entanglement is generated before the first teleportation, as a function of decoherence time before second teleportation. The first decoherence time yields $\varphi_1\tau = \pi/2$. Different curves correspond to different conditional evolution of the environment before the second teleportation with $|x|^2 = 0.1$ (dashed blue lines), $|x|^2 = 0.3$ (dashed-dotted red lines), and $|x|^2 = 0.5$ (solid black lines).

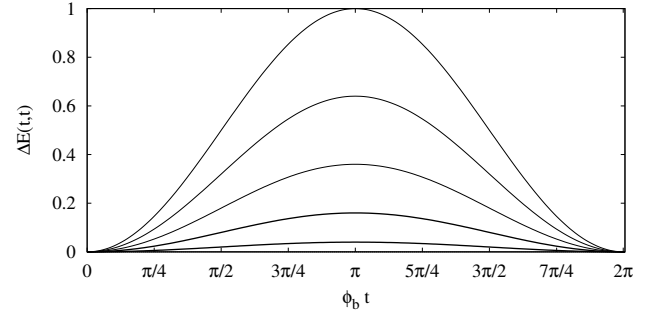

FIG. 2. Evolution of the difference between entanglement present in state (rhoaepsi) and state (rhoaephi) for a single qubit environment and exemplary pure-dephasing Hamiltonian with $x = y = 1/\sqrt{2}$. Different curves correspond to different initial state of the environment with $c_0 = 0.6, 0.7, 0.8, 0.9, 1$ going from the bottom curve to the top.

## VII. CONCLUSION

In this paper we have studied bidirectional teleportation of a qubit via a maximally entangled Bell state. We took into account a process leading to pure dephasing of the Bell state due to an interaction with an environment, but contrarily to previous works on the subject, we have kept the degrees of freedom of the environment throughout. Thanks to this, we were able to study the behavior of correlations with the environment while the teleportation procedure was operated.

Studying teleportation in one direction, we found that the correlations initially present between the Bell state and the environment are transported to the state of one qubit simultaneously as the unknown state is teleported to the same qubit. This yields a qubit-environment state which contains all information about the teleported qubit state, but the correlations with the environment mean that the state cannot be read out by measurements on the qubit alone (the information is not contained in the state of the qubit, but rather in the state of the qubit and environment).

If teleportation in the opposite direction is performed without additional decoherence, we observe that the state of the qubit and its correlations with the environment are teleported faithfully by the procedure. This means that the qubit-environment state pre- and post-teleportation is exactly the same, so also the state of the qubit alone was teleported with unit Fidelity. The two effects described above do not depend on the measurement outcome in the teleportation procedure.

If the second step of teleportation is preceded by a pure-dephasing process on the Bell state then the effect of teleportation does depend on the measurement outcome. In this case, due to correlations already present between one qubit and the environment, decoherence can be accompanied by the generation of entanglement between the three qubits and the environment, but also situations involving only generation of classical correlations are possible. During teleportation, the process of transport of correlations to only a single qubit and the teleportation of correlations between two qubits are intermixed. As a result there are two possible outcomes of teleportation, meaning that depending on the measurement outcome, to distinct qubit-environment states can be obtained. If the pre-teleportation qubit-environment state is separable then these two states must have the same amount of quantum correlations, although their degree of coherence can differ. If the pre-teleportation state is entangled then the states can not only differ in coherence, but also the amount of qubit-environment entanglement. The extreme case which can be realised has one of the teleported state contain only classical correlations while the other displays qubit-envronment entanglement.

[1] C. H. Bennett, G. Brassard, C. Crépeau, R. Jozsa, A. Peres, and W. K. Wootters, Teleporting an unknown quantum state via dual classical and einstein-podolsky-rosen channels, Phys. Rev. Lett. **70**, 1895 (1993).

[2] D. Gottesman and I. L. Chuang, Demonstrating the viability of universal quantum computation using teleportation and single-qubit operations, Nature **402**, 390 (1999).

[3] S. Bandyopadhyay and B. C. Sanders, Quantum teleportation of composite systems via mixed entangled states, Phys. Rev. A **74**, 032310 (2006).

[4] G.-F. Zhang, Thermal entanglement and teleportation in a two-qubit heisenberg chain with dzyaloshinski-moriya anisotropic antisymmetric interaction, Phys. Rev. A **75**, 034304 (2007).

[5] J.-C. Liu, Y.-H. Li, and Y.-Y. Nie, Controlled teleportation of an arbitrary two-particle pure or mixed state by using a five-qubit cluster state, International Journal of Theoretical Physics **49**, 1976–1984 (2010).

[6] P. Agrawal and A. K. Pati, Probabilistic quantum teleportation, Physics Letters A **305**, 12 (2002).

[7] F. Yan and T. Yan, Probabilistic teleportation via a non-maximally entangled ghz state, Chinese Science Bulletin **55**, 902–906 (2010).

[8] H. Prakash and V. Verma, Minimum assured fidelity and minimum average fidelity in quantum teleportation of single qubit using non-maximally entangled states, Quantum Information Processing **11**, 1951–1959 (2012).

[9] G. Bowen and S. Bose, Teleportation as a depolarizing quantum channel, relative entropy, and classical capacity, Phys. Rev. Lett. **87**, 267901 (2001).

[10] S. Oh, S. Lee, and H.-w. Lee, Fidelity of quantum teleportation through noisy channels, Phys. Rev. A **66**, 022316 (2002).

[11] F. Verstraete and H. Verschelde, Optimal teleportation with a mixed state of two qubits, Phys. Rev. Lett. **90**, 097901 (2003).

[12] M.-L. Hu, Environment-induced decay of teleportation fidelity of the one-qubit state, Physics Letters A **375**, 2140 (2011).

[13] K. Roszak and Ł. Cywiński, The relation between the quantum discord and quantum teleportation: The physical interpretation of the transition point between different quantum discord decay regimes, EPL (Europhysics Letters) **112**, 10002 (2015).

[14] L. Aolita, F. de Melo, and L. Davidovich, Open-system dynamics of entanglement:a key issues review, Reports on Progress in Physics **78**, 042001 (2015).

[15] D. Bouwmeester, J.-W. Pan, K. Mattle, M. Eibl, H. Weinfurter, and A. Zeilinger, Experimental quantum teleportation, Nature **390**, 575 (1997).

[16] D. Boschi, S. Branca, F. De Martini, L. Hardy, and S. Popescu, Experimental realization of teleporting an unknown pure quantum state via dual classical and einstein-podolsky-rosen channels, Phys. Rev. Lett. **80**, 1121 (1998).

[17] A. Furusawa, J. L. Sørensen, S. L. Braunstein, C. A. Fuchs, H. J. Kimble, and E. S. Polzik, Unconditional quantum teleportation, Science **282**, 706 (1998), https://science.sciencemag.org/content/282/5389/706.full.pdf.

[18] M. A. Nielsen, E. Knill, and R. Laflamme, Complete quantum teleportation using nuclear magnetic resonance, Nature **396**, 52 (1998).

[19] M. Riebe, H. Häffner, C. Roos, W. Hänsel, J. Benhelm, G. Lancaster, T. Körber, C. Becher, F. Schmidt-Kaler, D. James, et al., Deterministic quantum teleportation with atoms, Nature **429**, 734 (2004).

[20] M. Barrett, J. Chiaverini, T. Schaetz, J. Britton, W. Itano, J. Jost, E. Knill, C. Langer, D. Leibfried, R. Ozeri, et al., Deterministic quantum teleportation of atomic qubits, Nature **429**, 737 (2004).

[21] J. F. Sherson, H. Krauter, R. K. Olsson, B. Julsgaard, K. Hammerer, I. Cirac, and E. S. Polzik, Quantum teleportation between light and matter, Nature **443**, 557 (2006).

[22] H. Yonezawa, T. Aoki, and A. Furusawa, Demonstration of a quantum teleportation network for continuous variables, Nature **431**, 430 (2004).

[23] N. Takei, H. Yonezawa, T. Aoki, and A. Furusawa, High-fidelity teleportation beyond the no-cloning limit and entanglement swapping for continuous variables, Phys. Rev. Lett. **94**, 220502 (2005).

[24] R. Ursin, T. Jennewein, M. Aspelmeyer, R. Kaltenbaek, M. Lindenthal, P. Walther, and A. Zeilinger, Quantum teleportation across the danube, Nature **430**, 849 (2004).

[25] X.-S. Ma, T. Herbst, T. Scheidl, D. Wang, S. Kropatschek, W. Naylor, B. Wittmann, A. Mech, J. Kofler, E. Anisimova,

et al., Quantum teleportation over 143 kilometres using active feed-forward, Nature **489**, 269 (2012).

[26] J.-G. Ren, P. Xu, H.-L. Yong, L. Zhang, S.-K. Liao, J. Yin, W.-Y. Liu, W.-Q. Cai, M. Yang, L. Li, K.-X. Yang, X. Han, Y.-Q. Yao, J. Li, H.-Y. Wu, S. Wan, L. Liu, D.-Q. Liu, Y.-W. Kuang, Z.-P. He, P. Shang, C. Guo, R.-H. Zheng, K. Tian, Z.-C. Zhu, N.-L. Liu, C.-Y. Lu, R. Shu, Y.-A. Chen, C.-Z. Peng, J.-Y. Wang, and J.-W. Pan, Ground-to-satellite quantum teleportation, Nature **549**, 70 (2017).

[27] Y. Yeo, Local noise can enhance two-qubit teleportation, Phys. Rev. A **78**, 022334 (2008).

[28] E.-M. Laine, H.-P. Breuer, and J. Piilo, Nonlocal memory effects allow perfect teleportation with mixed states, Scientific Reports **4**, 4620 (2014).

[29] L. Vaidman, Teleportation of quantum states, Phys. Rev. A **49**, 1473 (1994).

[30] E. O. Kiktenko, A. A. Popov, and A. K. Fedorov, Bidirectional imperfect quantum teleportation with a single bell state, Phys. Rev. A **93**, 062305 (2016).

[31] K. Roszak, Criteria for system-environment entanglement generation for systems of any size in pure-dephasing evolutions, Phys. Rev. A **98**, 052344 (2018).

[32] B.-G. Englert and J. A. Bergou, Quantitative quantum erasure1dedicated to marlan scully on the occasion of his 60th birthday.1, Optics Communications **179**, 337 (2000).

[33] K. Roszak and P. Machnikowski, "Which path" decoherence in quantum dot experiments, **351**, 251 (2006).

[34] K. Roszak and L. Cywiński, Characterization and measurement of qubit-environment-entanglement generation during pure dephasing, Phys. Rev. A **92**, 032310 (2015).

[35] K. Roszak, Measure of qubit-environment entanglement for pure dephasing evolutions, Phys. Rev. Research **2**, 043062 (2020).

[36] K. Roszak and L. Cywiński, Equivalence of qubit-environment entanglement and discord generation via pure dephasing interactions and the resulting consequences, Phys. Rev. A **97**, 012306 (2018).

[37] K. Roszak, D. Kwiatkowski, and L. Cywiński, How to detect qubit-environment entanglement generated during qubit dephasing, Phys. Rev. A **100**, 022318 (2019).

[38] B. Rzepkowski and K. Roszak, A scheme for direct detection of qubit-environment entanglement generated during qubit pure dephasing (2020), arXiv:2002.10901 [quant-ph].

[39] K. Roszak and L. Cywiński, Qubit-environment-entanglement generation and the spin echo, Phys. Rev. A **103**, 032208 (2021).